\documentclass[12pt]{article}
\usepackage{epsf}
\def\beq{\begin{equation}}
\def\eeq{\end{equation}}
\def\ap#1#2#3 {Ann. Phys. (NY) {\bf#1} (19#2) #3}
\def\err#1#2#3 {{\it Erratum} {\bf#1} (19#2) #3}
\def\ib#1#2#3 {{\it ibid.} {\bf#1} (19#2) #3}
\def\ijmp#1#2#3 {Int. J. Mod. Phys. {\bf#1} (19#2) #3}
\def\jetp#1#2#3 {JETP Lett. {\bf#1} (19#2) #3}
\def\mpl#1#2#3 {Mod. Phys. Lett. {\bf#1} (19#2) #3}
\def\np#1#2#3 {Nucl. Phys. {\bf#1} (19#2) #3}
\def\pl#1#2#3 {Phys. Lett. {\bf#1} (19#2) #3}
\def\prep#1#2#3 {Phys. Rep. {\bf#1} (19#2) #3}
\def\prev#1#2#3 {Phys. Rev. {\bf#1} (19#2) #3}
\def\prl#1#2#3 {Phys. Rev. Lett. {\bf#1} (19#2) #3}
\def\sjnp#1#2#3 {Sov. J. Nucl. Phys. {\bf#1} (19#2) #3}
\def\spj#1#2#3 {Sov. Phys. JETP {\bf#1} (19#2) #3}
\def\spu#1#2#3 {Sov. Phys. Usp. {\bf#1} (19#2) #3}
\def\zp#1#2#3 {Zeit. Phys. {\bf#1} (19#2) #3}

\begin{document}
\begin{titlepage}
\begin{center}
{\Large \bf William I. Fine Theoretical Physics Institute \\
University of Minnesota \\}  \end{center}
\vspace{0.2in}
\begin{flushright}
FTPI-MINN-04/05 \\
UMN-TH-2232-04 \\
January 2004 \\
\end{flushright}
\vspace{0.3in}
\begin{center}
{\Large \bf  Quarkonium chromo-polarizability from the decays
$J/\psi(\Upsilon) \to \pi \pi \ell^+ \ell^-$
\\}
\vspace{0.2in}
{\bf M.B. Voloshin  \\ }
William I. Fine Theoretical Physics Institute, University of
Minnesota,\\ Minneapolis, MN 55455 \\
and \\
Institute of Theoretical and Experimental Physics, Moscow, 117259
\\[0.2in]
\end{center}

\begin{abstract}
It is pointed out that the diagonal amplitude of the $E1-E1$
chromo-electric interaction with soft gluon fields
(chromo-polarizability) can be measured directly for the $J/\psi$ and
$\Upsilon$ resonances in the decays $J/\psi \to \pi \pi \ell^+ \ell^-$
and $\Upsilon \to \pi \pi \ell^+ \ell^-$ with soft pions. For the
$J/\psi$ this amplitude is often discussed in connection with the
$J/\psi$ interaction with nuclear matter, while for the $\Upsilon$ the
chromo-polarizability enters the estimates of the non-perturbative mass
shift of the resonance relevant to precision determination of the $b$
quark mass from the $\Upsilon$ mass.
\end{abstract}

\end{titlepage}

It is realistic to expect that a very high statistics data on heavy
quarkonia will become available in a foreseeable future. In particular
the plans of the CLEO-c experiment\cite{cleoc} include acquiring about
$10^9$ events at the $J/\psi$ resonance and, given the capabilities of
the $B$ factories, a great increase in the statistics of the $\Upsilon$
resonances is also quite possible. With such amount of data quite rare
processes can be studied, which can help in resolving some of the
old-standing problems in dynamics of heavy quarkonia. The purpose of the
present letter is to point out that a study of the decays  $J/\psi \to
\pi \pi \ell^+ \ell^-$ and $\Upsilon \to \pi \pi \ell^+ \ell^-$ with
soft pions would allow to measure the gluon polarizability of the
respective quarkonium resonance, which determines the strength of the
quarkonium interaction with soft gluon field, and which thus far is only
guessed theoretically.

The chromo-polarizability $\alpha$ of a state of quarkonium can be
defined (in complete analogy with the usual polarizability of atoms)
through the effective Hamiltonian of the quarkonium state interaction
with a soft chromo-electric field ${\vec E}^a$:
\beq
H_{eff}=-{1 \over 2} \, \alpha \, {\vec E}^a \cdot {\vec E}^a~.
\label{heff}
\eeq
This effective interaction arises in the second order in the leading
$E1$ chromo-electric dipole term in the multipole
expansion\cite{gottfried, mv} of the interaction of a heavy quarkonium
with the gluon field:
\beq
H_{E1}=-{1 \over 2} \xi^a \, {\vec r} \cdot {\vec E}^a (0)~,
\label{e1}
\eeq
where $\xi^a=t_1^a-t_2^a$ is the difference of the color generators
acting on the quark and antiquark (e.g. $t_1^a = \lambda^a/2$ with
$\lambda^a$ being the Gell-Mann matrices), and ${\vec r}$ is the vector
for relative position of the quark and the antiquark. (In the
normalization used throughout this paper the QCD coupling $g$ is
included in the definition of the field, so that e.g. the gluon field
Lagrangean reads as $L=-(F_{\mu \nu}^a)^2/(4 g^2)$.) Thus for a
colorless $S$ wave state the chromo-polarizability $\alpha$ is given by
the diagonal matrix element:
\beq
\alpha={1 \over 48} \,\langle S |\, \xi^a \, r_i \, G_A \, r_i \, \xi^a 
| S \rangle ~,
\label{dme}
\eeq
where $G_A$ is the Green's function for a heavy quark pair in
color-octet (adjoint) state. A theoretical understanding of such matrix
element is at least highly model dependent at present due to its
sensitivity to the wave function of the quarkonium and due to presently
unknown propagator $G_A$ of a colored quark pair.

The chromo-polarizability of $J/\psi$: $\alpha_{J/\psi}$, enters as an
important parameter in analyses of the $J/\psi$ interaction with nuclear
matter (see e.g. in Refs. \cite{kv,kharzeev}) and is estimated
essentially on dimensional grounds\cite{bp}. The same quantity for the
$\Upsilon$ resonance: $\alpha_\Upsilon$, determines\cite{mv} the
non-perturbative shift of the $\Upsilon$ mass due to the gluon vacuum
condensate\cite{svvz} $\langle 0 | (F_{\mu \nu}^a)^2 | 0 \rangle$:
\beq
\delta M_\Upsilon = {1 \over 8} \, \alpha_\Upsilon \, \langle 0 |
(F_{\mu \nu}^a)^2 | 0 \rangle~.
\label{dmy}
\eeq

The knowledge of the non-perturbative shift of the $\Upsilon$ mass is
important for the method of determining the $b$ quark mass by comparing
$M_\Upsilon$ with the perturbative QCD expression in terms of $m_b$ for
the mass of the lowest $^3S_1$ bound state of the quark-antiquark pair
(for a detailed discussion see the review \cite{ekl} and references
therein). Practical estimates of this non-perturbative shift use an
extrapolation down to the bottomonium of the
formulas\cite{leutwyler,mv2} for asymptotically heavy quarkonium. In the
latter limit the heavy quarkonium is essentially a Coulomb system, and
the matrix elements of the type in eq.(\ref{dme}) can be found
explicitly. In particular for the $1S$ state the chromo-polarizability
is found\cite{leutwyler,mv2} as
\beq
\alpha_{1S}={78 \over 425} \, {m_Q \over k_B^4}~,
\label{acoul}
\eeq
with $k_B=2 m_Q \alpha_s(k_B)/3$ being the Bohr momentum for the heavy
quarkonium. If applied to bottomonium, this formula gives
$\alpha_\Upsilon \approx 1 GeV^{-3}$. Using then the value of the gluon
vacuum condensate\cite{svvz} $\langle 0 | (F_{\mu \nu}^a)^2 | 0
\rangle=\langle 0 |4 \pi \alpha_s \, (G_{\mu \nu}^a)^2 | 0 \rangle
\approx 0.5 \, GeV^4$, one estimates from eq.(\ref{dmy}) the
non-perturbative shift of the mass of $\Upsilon$: $\delta M_\Upsilon
\approx 50 - 60 \, MeV$. However, in addition to the uncertainty in the
value of the gluon condensate, there are at least two other
uncertainties involved in this estimate: one arising from the
application of the asymptotic expression (\ref{acoul}) to bottomonium,
and the other associated with possible contribution of vacuum averages
of higher dimension, generally nonlocal, gluonic operators. Although it
is not clear at present to what extent the latter uncertainty can be
estimated, a direct measurement of the chromo-polarizability
$\alpha_\Upsilon$ would definitely fix at least one factor in this
problem.

It should be noted that the non-diagonal amplitudes of the type in
eq.(\ref{dme}) for the transitions between the $2S$ and $1S$ states,
\beq
\alpha_{1S-2S}={1 \over 48} \,\langle 1S |\, \xi^a \, r_i \, G_A \, r_i
\, \xi^a  | 2S \rangle ~,
\label{ndme}
\eeq
can be found both in charmonium and bottomonium from the spectra and the
rates of the pionic transitions $\psi^\prime \to \pi \pi \, J/\psi$ and
$\Upsilon^\prime \to \pi \pi \, \Upsilon$. The relation arises through
the fact that the dominant part of the amplitude of production of two
pions by the gluonic operator $({\vec E}^a)^2$ is determined\cite{vz} by
the trace anomaly in QCD and the chiral algebra:
\beq
\langle \pi^+ \pi^- | \,  ({\vec E}^a)^2 | 0 \rangle = {8 \pi^2 \over b}
q^2 + O(\alpha_s q_0^2) + O(m_\pi^2)~,
\label{pime}
\eeq
where $q=p_++p_-$ is the total 4-momentum of the pion pair (so that
$q_0$ is the total energy of the pair), $b=9$ is the first coefficient
in the QCD beta function with three light quarks, and the subleading
terms, analyzed in Ref.\cite{ns}, are relatively slowly varying with
$q^2$ in the physical region of the pionic transition. In practice these
terms can be approximated by a constant $C$, which depends on $q_0$, so
that the amplitude of the decay can be written according to the
equations (\ref{e1}), (\ref{ndme}), and (\ref{pime}) as
\beq
A(2S \to \pi^+ \pi^- \, 1S) = -{4 \pi^2 \over b} \, \alpha_{1S-2S} \,
(q^2-C)
\label{a2pi}
\eeq
with $C$ being slightly different for the transitions in charmonium and
in bottomonium. The observed experimental spectra of the dipion mass in
the decays $\psi^\prime \to \pi \pi \, J/\psi$ and $\Upsilon^\prime \to
\pi \pi \, \Upsilon$ agree with the form of the amplitude in
eq.(\ref{a2pi}) and the with the numerical value of the constant: $C =
(4.6 \pm 0.2)\, m_\pi^2$ for the transition in charmonium and $C = (3.3
\pm 0.2) \, m_\pi^2$ for the decay $\Upsilon^\prime \to \pi \pi \,
\Upsilon$ (for a discussion see e.g. the review \cite{vzai}).  

The total rates of the pionic transitions are calculated from
eq.(\ref{a2pi}) as
\begin{eqnarray}
&&\Gamma(\psi^\prime \to \pi^+ \pi^- \, J/\psi) \approx 0.30 \, {8 \pi
\over 105 \, b^2} \, |\alpha_{J/\psi - \psi^\prime}|^2 \, \left [
M(\psi^\prime)- M(J/\psi) \right]^7~,
\nonumber \\
&&\Gamma(\Upsilon^\prime \to \pi^+ \pi^- \, \Upsilon) \approx 0.36 \, {8
\pi \over 105 \, b^2} \, |\alpha_{\Upsilon - \Upsilon^\prime}|^2 \,
\left [ M(\Upsilon^\prime)- M(\Upsilon) \right]^7~,
\label{num}
\end{eqnarray}
where the decimal numerical factors describe the relative suppression of
the rates due to the corresponding constant term $C$ and due to the
nonzero pion mass. Thus from a comparison of these expressions with the
experimental data\cite{pdg} on the rates one estimates the transition
chromo-polarizabilities:
\beq
|\alpha_{J/\psi - \psi^\prime}| \approx 2.0 \, GeV^{-3}~,
~~~|\alpha_{\Upsilon - \Upsilon^\prime}| \approx 0.66 \, GeV^{-3}~.
\label{eat}
\eeq
These estimates of the transition matrix elements illustrate the typical
values that one can expect for the diagonal chromo-polarizability in
charmonium and bottomonium. It is also somewhat satisfying to notice
that for the bottomonium the estimate of the diagonal
chromo-polarizability from eq.(\ref{acoul}) is in a reasonable agreement
with the value of the transition term, since one generally would expect
the diagonal matrix element (\ref{dme}) to be somewhat larger than the
non-diagonal (\ref{ndme}).

Proceeding to discussion of the decays $J/\psi \to \pi \pi \ell^+
\ell^-$ and $\Upsilon \to \pi \pi \ell^+ \ell^-$ we retain the notation
$q$ for the total 4-momentum of the two pions. The amplitude of such
decay for a $1^3S_1$ state of heavy quarkonium in the soft pion limit
can be written as a sum over intermediate $n^3S_1$ states:
\beq
A(1^3S_1 \to \pi^+ \pi^- \ell^+ \ell^-) = {1 \over 2} \, \langle \pi^+
\pi^- | \,  ({\vec E}^a)^2 | 0 \rangle \, \sum_{n=1} \, {\alpha_{1S -
nS} \over M(nS)-M(1S)+q_0 } \, A(n^3S_1 \to \ell^+ \ell^-)~,  
\label{adec}
\eeq
where the sum goes over the discrete states as well as the continuum. In
writing this expression it is taken into account that the soft pion
approximation is only valid at $q_0 \ll M(1S)$, so that any recoil of
the heavy quarkonium upon emission of the pion pair can be and is
neglected. In this limit the relation between $q_0$ and the total
momentum $l$ of the lepton pair can also be written as $M^2(1S)-l^2=2 \,
q_0 \, M(1S)$.

In the chiral limit the first term (with $n=1$) in the sum in
eq.(\ref{adec}) dominates for soft pions, due to its singular behavior
as $1/q_0$. It can be noticed however that the decay amplitude itself is
not singular due to the (even faster) vanishing of the pion production
amplitude (\ref{pime}) in the limit of soft pions. In the `real life'
the minimal practical energy $q_0$ is not much less than the spacing of
the quarkonium levels, and the contribution of higher states mixes into
the amplitude. In what follows we first consider the contribution of
only the first term of the sum in eq.(\ref{adec}) and then discuss the
effect of the higher terms. Keeping only the contribution of the first
term in the sum in eq.(\ref{adec}) one can write the differential rate
of the discussed decay in terms of the chromo-polarizability
$\alpha_{1S}$ and the leptonic width $\Gamma_{ee}(1^3S1) \equiv
\Gamma(1^3S_1 \to \ell^+ \ell^-)$ in the form
\beq
d \Gamma(1^3S_1 \to \pi^+ \pi^- \ell^+ \ell^-) = {[q^2-C(q_0)]^2 \over 4
b^2 \, q_0^2} \, |\alpha_{1S}|^2  \, \sqrt{1-{4 m_\pi^2 \over q^2}} \,
\sqrt{q_0^2-q^2} \, \Gamma_{ee}(1^3S1) \, d q^2 \, d q_0 ~,
\label{dgam}
\eeq
where the formula (\ref{pime}) is used with a simplified parametrization
of the subleading terms as a constant $C(q_0)$ similar to that in
eq.(\ref{a2pi}).

In order to assess the feasibility of observing the discussed decays it
can be noted that at a given constraint on the maximal value of $q_0$:
$q_0 < \Delta$ (or equivalently at a lower cutoff on the invariant mass
of the lepton pair) the probability described by eq.(\ref{dgam})
strongly peaks near the highest values of both $q^2$ and $q_0$, i.e $q^2
\sim \Delta^2$ and $q_0 \sim \Delta$, and the total probability in the
kinematical region constrained as $q_0 < \Delta$ scales approximately as
$\Delta^6$. However at higher $q^2$ both the dominance of the diagonal
$1S - 1S$ transition in the sum in eq.(\ref{adec}) becomes weaker and
the linear in $q^2$ behavior of the amplitude in eq.(\ref{pime}) derived
for soft pions becomes questionable. It is still quite likely that with
these limitations the presented here approach can be used up to somewhat
higher values of $\Delta$: $\Delta \approx 0.8 - 0.9 \, GeV$, than those
observed in the pionic transitions from $\psi^\prime$ and
$\Upsilon^\prime$. Indeed, experimentally the linearity in $q^2$ of the
amplitude of the transition $\Upsilon^\prime \to \pi \pi \, \Upsilon$ is
very accurate\cite{argus,vzai} in the physical region, i.e. up to
$\sqrt{q^2} \approx 0.56 \, GeV$ with no obvious hint of its violation
close to this region. On the other hand the $f_0(980)$ resonance places
a natural upper bound on the region of applicability of eq.(\ref{pime}).
As to the contribution of higher quarkonium states in the sum in
eq.(\ref{adec}), for each of these states the magnitude of this
contribution relative to that of the diagonal transition is given  by
\beq
r_n = \left | {\alpha_{1S-nS} \over \alpha_{1S}} \, {q_0 \over
M(n^3S_1)-M(1^3S_1)+q_0} \right | \, \left [ {\Gamma_{ee}(n^3S_1) \over
\Gamma_{ee}(1^3S_1)} \right ]^{1/2}~.
\label{rn}
\eeq
The transition polarizability $\alpha_{1S-nS}$ should considerably
decrease with $n$. This is supported by the very small experimental rate
of the decay $\Upsilon(3S) \to \pi \pi \, \Upsilon$\footnote{The known
problem with this decay is that the observed dipion mass spectrum does
not agree with the formula in eq.(\ref{pime}) (see e.g. in the review
\cite{vzai}). A natural although quite qualitative explanation of this
fact is that the polarizability $\alpha_{1S-3S}$ is very small, and
subleading effects in the multipole expansion in QCD come into play in
this transition.}. Thus, most likely, the only real effect of higher
states up to $\Delta \sim 0.9 \, GeV$ reduces to that of the 2S
resonances. These however can be accounted for in the data analysis,
since for these resonances all the parameters (except for the overall
relative phase of their contribution) in eq.(\ref{rn}) are known.
Furthermore an observation and analysis of the discussed decays at
higher values of $q_0$, where the expression (\ref{pime}) is no longer
valid, would be of a great interest for studies of the pion-pion
scattering beyond the soft-pion region.

Numerically one can estimate from eq.(\ref{dgam}) the total rate in the
kinematical region constrained by $\Delta = 0.9 \, GeV$ as
\beq
\Gamma(1^3S_1 \to \pi^+ \pi^- \ell^+ \ell^-)|_{q_0 < 0.9 \, GeV} \approx
10^{-4} \left | {\alpha_{1S} \over 2 \, GeV^{-3}} \right |^2 \,
\Gamma_{ee}(1^3S_1)~.
\label{numres}
\eeq
Given that the diagonal polarizability is likely to be larger than the
transition one, it can be expected that for the $J/\psi$ resonance the
branching ratio of the discussed decay in a useable kinematical range
should be at the level of $10^{-5}$ which looks to be well within the
reach with the expected CLEO-c data sample. For the $\Upsilon$
resonances the effect is reduced by a factor of about 20 due to a
smaller chromo-polarizability (cf. eq.(\ref{eat})) and also due to a
smaller value of $B(\Upsilon \to \ell^+ \ell^-)$. Thus an experimental
study of the discussed decay for the $\Upsilon$ resonance is likely to
be a future task for a $10^{35}$ upgrade of KEKB\cite{1035} and/or the
$10^{36}$ B-factory\cite{1036}.

\section*{Acknowledgments}
This work is supported in part by the DOE grant DE-FG02-94ER40823.

\end{document}